# Spin ordering of nuclear spectra from random interactions [*]

D. Mulhall, V. Zelevinsky, and A. Volya

National Superconducting Cyclotron Laboratory, Michigan State University, East Lansing, Michigan 48824-1321

Recent developments in many-body quantum chaos have raised the issue of correlations between different families of levels in the spectra of random fermionic systems. It seems that rotational invariance is sufficient to force an otherwise random system to have statistically ordered spectra where the ground state can have zero spin or the maximum spin allowed a huge portion of the time. We have tackled this question and have characterized the role of "geometric chaoticity" and pairing in random systems. We have shown how the geometry gives rise to such a distribution of ground state spins and examined the coexistence of these regularities with chaotic wave functions.

## 1. Introduction

The role of chaotic dynamics in nuclear physics has been extensively studied. Geometric chaoticity, or the assumption of pseudorandom coupling of angular momentum, was used to derive the early level density formulae. Random matrix theory clarified the role of global symmetries and lead to new methods of statistical analysis of nuclear spectra, while work with shell model wave functions gave a deeper understanding on the thermodynamics of the nucleus. The question of correlations between different classes of states described by the same hamiltonian has been neglected. Recent investigations [1, 2] showed that the spectra of random hamiltonians which respected global symmetries exhibited strong correlations between the different classes. This phenomenon was subsequently observed in a variety of systems but recently a reasonable physical explaination was proposed [3]. This work addresses this important issue and explains the correlations in terms of geometric chaoticity. The preference of the systems under study to have spin zero or maximum spin ground states and parabolic yrast lines is shown to have its origins in the geometry of the system, as opposed to the specific details of the interaction, i.e. it is a direct result of the complex coupling of many single-particle angular momenta required to build a many-body state with the appropriate symmetry.

---

[*] Presented at "High Spin Physics 2001", University of Warsaw, Poland, February 2001.

In order to address the regularities in the spectra of random systems we choose to study the simplest possible system, that of $N$ fermions on a single energy level of spin $j$ which has a $2j+1$ degeneracy. The regularities are robust with respect to the choice of both system and ensemble. Within this space, the general two-fermion rotationally invariant Hamiltonian can be written as

$$H = \sum_{L\Lambda} V_L P^\dagger_{L\Lambda} P_{L\Lambda}, \qquad (1)$$

where the operators for a pair of fermions with total angular momentum $L$ and projection $\Lambda$ are defined as

$$P^\dagger_{L\Lambda} = \frac{1}{\sqrt{2}} \sum_{mn} C^{L\Lambda}_{mn} a^\dagger_m a^\dagger_n, \quad P_{L\Lambda} = \frac{1}{\sqrt{2}} \sum_{mn} C^{L\Lambda}_{mn} a_n a_m; \qquad (2)$$

$a_m$ annihilates a particle with single particle angular momentum $j$ and projection $m$, $C^{L\Lambda}_{mn}$ are the Clebsch-Gordan coefficients $\langle L\Lambda | jm; jn \rangle$, and the two-particle states $|2; L\Lambda\rangle = P^\dagger_{L\Lambda}|0\rangle$ are properly normalized. Because of Fermi statistics, only even values of $L$ are allowed in the single-$j$ space. There are $j + \frac{1}{2}$ interaction parameters $V_L$ and these numbers define our ensemble. The ensemble we chose was a uniform distribution of $V_L$ between -1 and 1, which sets the energy scale. We analyzed many realizations of our ensemble for various values of particle number, $N$, and single-particle angular momentum $j$. Spectra were obtained by direct diagonalization of the hamiltonian matrix for a given set of $V_L$ but as the dimensionality got too large the shell model code OXBASH was used to get the spectra and wave functions.

In the $N = 4$ and 6 particle cases there was a huge fraction of systems with ground state spin $J_0 = 0$ and $J_0 = J_{max}$ where $J_{max}$ is the maximum allowed value of angular momentum. We will call these fractions $f_0$ and $f_{Jmax}$ respectively, see Fig. 1. The predominance of these values of angular momentum occurs in spite of the fact that they account for a small fraction of the total states. In the $N = 5$ particle systems there was a large fraction $f_j$ of ground states with total spin $J = j$ which is consistent with the picture of a valence spin $j$ particle on a spin 0 core.

The mean yrast line for all spectra with a spin 0 ground state was a parabolic function of $J$ with a strong odd-even effect, see Fig. 2. For the maximum spin ground state spectra the parabola is inverted. These smooth yrast lines show a strong correlation between levels with different quantum numbers. These correlations coexist with the familiar ones of random matrix theory. The distribution $P(s)$ of nearest neighbor level spacings within a set of levels with the same angular momentum was in full agreement with the Wigner distribution.

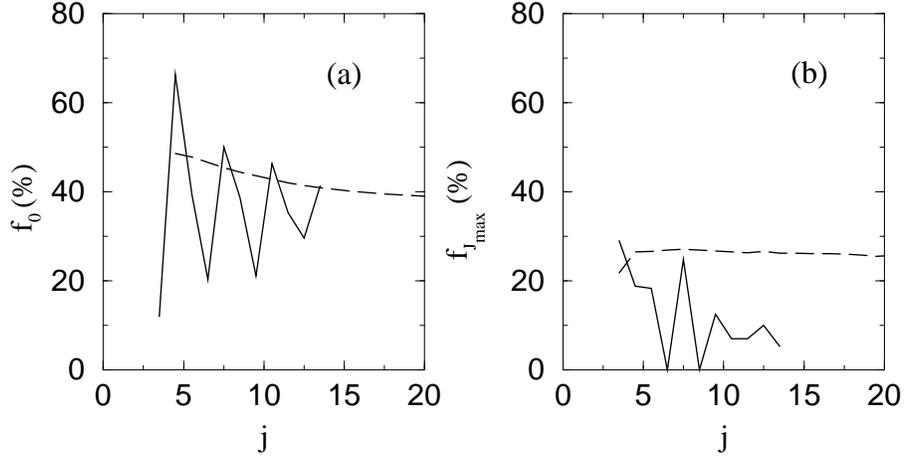

Fig. 1. The fraction of ground states with $J = 0$ for $N = 4$ and different $j$. Ensemble results; solid line. Statistical theory; dotted line, panel (a). Upper limit for $f_{J_{max}}$ from the statistical theory and analysis of the $J_{max}$ region; dotted line, panel (b).

## 2. The Statistical Model

The system was treated with an approach based on equilibrium statistical mechanics which leads to the Fermi-Dirac distribution

$$n_m = \frac{1}{\exp(\gamma m - \mu) + 1} \quad (3)$$

under the constraints

$$N = \sum_m n_m, \quad M = \sum_m m n_m. \quad (4)$$

The quantities $\mu(N, M)$, the chemical potential, and $\gamma(N, M)$, the cranking frequency, are the Lagrange multipliers associated with the two constraints.

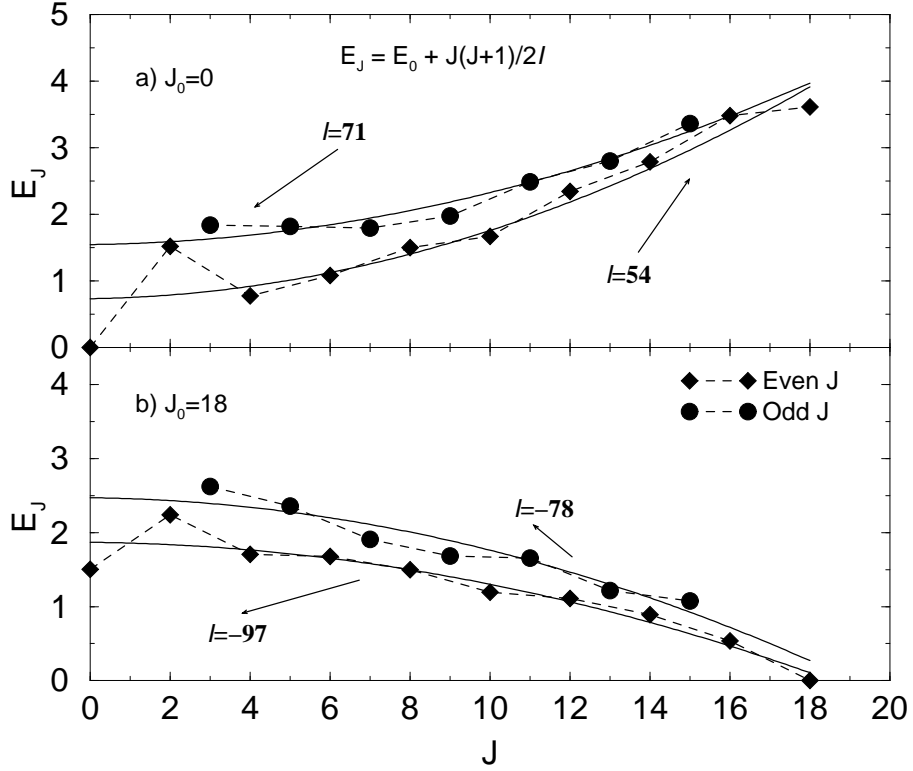

Fig. 2. The mean lowest energies vs $J$ for the $N = 6$, $J = 11/2$ ensemble for spectra with (a) $J_0 = 0$ and (b) $J_0 = J_{max}$. Separate values of the moment of inertia $I$ were extracted from the odd and even $J$ points..

The equilibrium energy for an $N$ particle-state with total projection $M$ is

$$\langle H \rangle = \sum_{L,\Lambda,1,2} V_L \left| C_{12}^{L\Lambda} \right|^2 \langle n_1 n_2 \rangle. \tag{5}$$

Treating $\gamma$ as an expansion parameter, expanding $n_m$, performing the summations over the various products of single particle projections and Clebsch-Gordan coefficients, and identifying $M$ with $J$ we come to

$$\langle H \rangle_{N,J} = \sum_L (2L+1) V_L \left[ h_0(L) + h_2(L) J^2 + h_4(L) J^4 \right] \tag{6}$$

where, with $\Omega = 2j + 1$,

$$h_0(L) = \frac{N^2}{\Omega^2}, \quad h_2(L) = \frac{3}{2\mathbf{j}^4 \Omega^2} (\mathbf{L}^2 - 2\mathbf{j}^2),$$

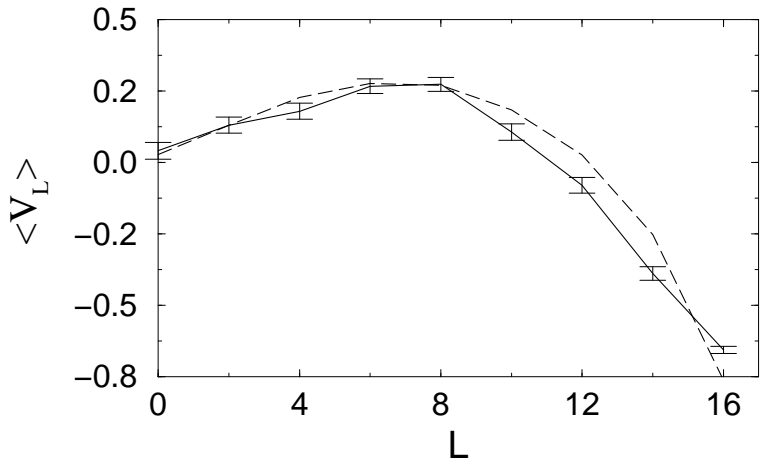

Fig. 3. The average values of the random parameters $V_L$ for the ground state spin $J_0 = J_{max}$, $N = 4$, and $j = 17/2$ as a function of $L$ shown with the statistical errors (solid line), and the results of the statistical model for $h_2 < 0$ (dashed line).

$$h_4(L) = \frac{9(\Omega - 2N)^2}{40\,\mathbf{j}^8(\Omega - N)^2 N^2 \Omega^2}\left(3\mathbf{L}^4 + 3\mathbf{L}^2 - 12\mathbf{j}^2\mathbf{L}^2 - 6\mathbf{j}^2 + 8\mathbf{j}^4\right). \quad (7)$$

The regularities of the ensemble are insensitive to the choice of distribution of interaction parameters $V_L$ as long as the choice is symmetric about zero. The distribution of the sum in Eq. (6) can be calculated assuming a Gaussian distribution for each $V_L$. Our model captures the upper limit for $f_0$ in the systems of various $j$, Fig. 1. The staggering of $f_0$ in the numerical results for $N = 4$, with its period of 3 in $j$, can be understood by observing the behaviour of the multiplicity of the spin 0 states as $j$ changes. It is incremented by 1 for every time $j$ is incremented by 3. The theory overestimates the value of $f_{Jmax}$ but the expansion of the $\gamma$ parameter was valid only for small values of $M$ and thus $J$.

We can use (6) to extract the mean values $\langle V_L \rangle$ of those sets of interaction parameters that generate spectra with certain $J_0$. In Fig. 3 we compared the theoretical points for $\langle V_L \rangle$ vs. $L$ for $N = 6$ systems with the actual mean values of random parameters $V_L$ which led to $J_0 = J_{max}$. The resulting curve is in total agreement with our statistical theory.

The equilibrium energies from Eq. (6) are highly correlated with the numerical values, differing by a constant negative shift. When plotted against each other the points fell along a straight line with a slope of $0.81 \pm 0.02$ and $0.81 \pm 0.03$ for those systems with $J_0 = 0$ and 30, respectively, for all values of $J$, see Fig. 4. There was an odd-even staggering in the value of the negative shift and its magnitude was smaller for $J_0 = 30$, Fig. 5.

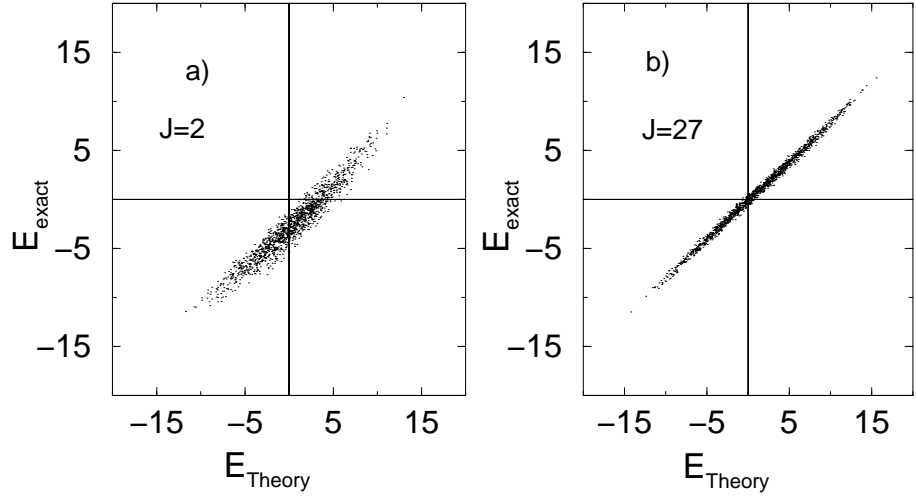

Fig. 4. Numerical values of the lowest $J = 2$ (a), and $J = 27$ (b), energies vs. predictions of the statistical model, eq. (6), for the $J_0 = 0$ spectra of the $N = 6$, $j = 15/2$ ensemble.

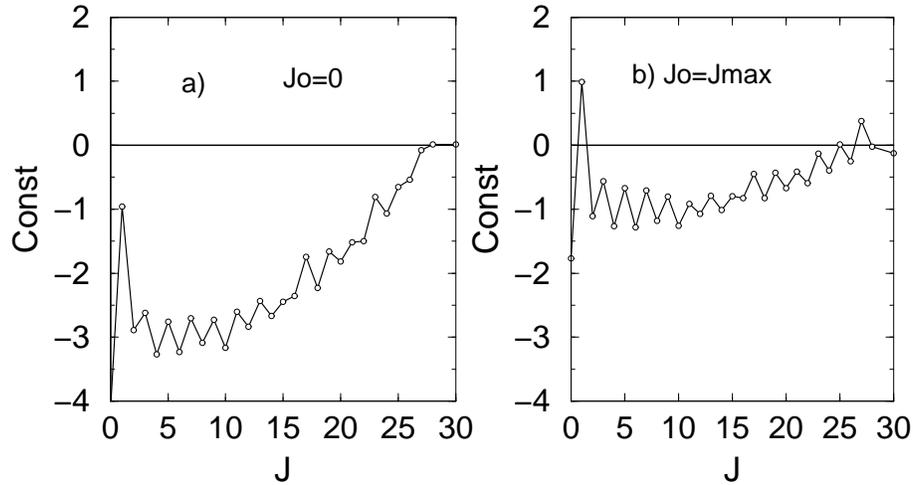

Fig. 5. The constant shift in the theoretical equilribium energies of spin $J$ vs. $J$ for the $N = 6$ $j = 15/2$ ensemble for the two values of ground state spin.

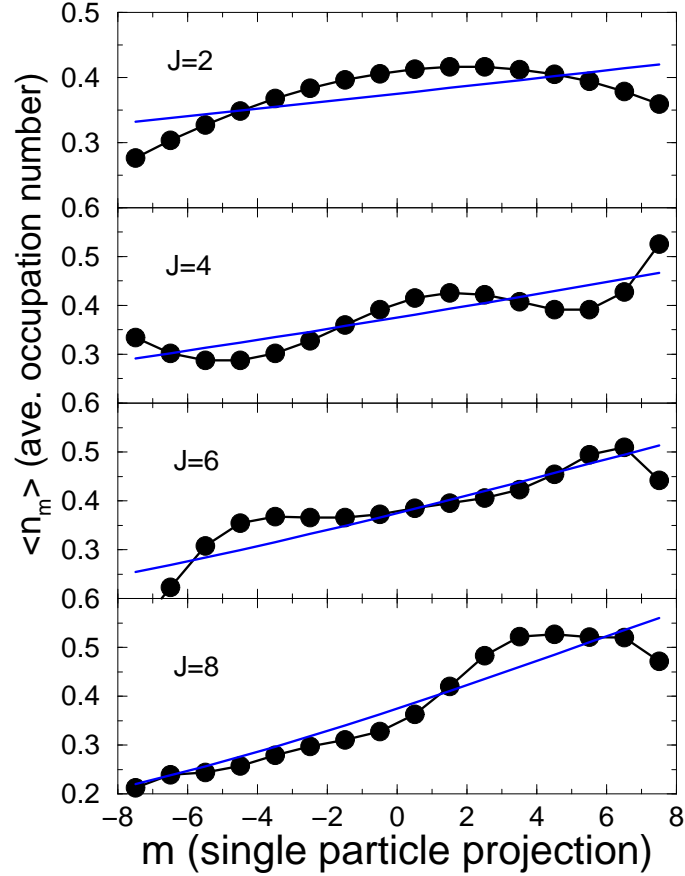

Fig. 6. The average occupation numbers for the 1417 out of 2000 (70%) spectra with $J_0 = 0$ for the $N = 6$, $j = 15/2$ ensemble.

The assumption of statistical equilribium implies that the off-diagonal matrix elements of $a_m$ and $a_m^\dagger$ and their products make an incoherent contribution to (6). We also assumed that the occupation numbers are uncorrelated. Actually using the numerical results for the correlated occupation numbers in (5) resulted in a slope close to unity in the plots of Fig. (4). In Fig. 6 we see the average values of $n_m$ for the lowest $J = 2, 4, 6, 8$ energy levels of those $N = 6$, $j = 15/2$ spectra with $J_0 = 0$. They oscillate about the statistical values, exhibiting a shell structure that the theory smeared out through the approximation of uncorrelated occupancies.

In spite of the regularities of the ground state spin, their origin in random spin coupling leads us to expect the corresponding wave functions to be chaotic. As a reference point, we look at the distribution of overlaps, $x$, between the ground state wave functions of our ensemble and the spin zero seniority zero wavefuntion which appears as the ground state for the regular pairing interaction, $V_0 = -1$, $V_{L\neq 0} = 0$,

$$x = |\langle J = 0, \text{g.s.}|0, \text{p}\rangle|^2. \tag{8}$$

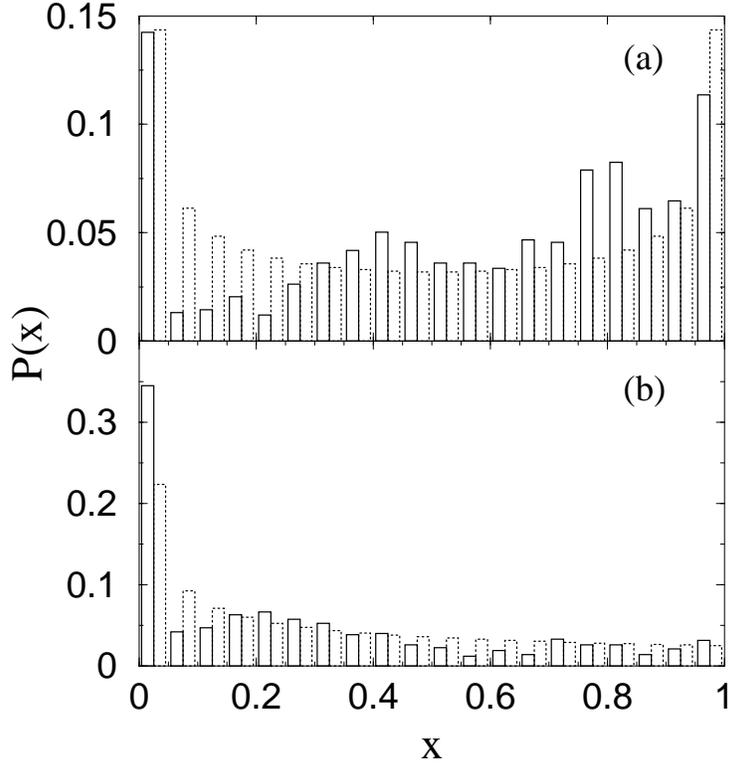

Fig. 7. The distribution of overlaps, eq. (8), of ground states with $J_0 = 0$ with the fully paired (seniority zero) state, solid histograms, for the ensemble with all $V_L$ random, panel (b), and for the ensemble with random $V_{L\neq 0}$ and regular pairing, $V_{L=0} = -1$, panel (a). Dashed histograms show the predictions for chaotic wave functions of dimension $d = 3$, panel (b), and $d = 2$, panel (a).

The histogram of the distribution $\mathcal{P}(x)$ is presented in Fig. 7(b) for the case of $N = 6$ particles on the $j = 11/2$ level where there are 3 states of spin

mulhall   printed on November 19, 2018   9$J = 0$. In the chaotic limit this would lead to

$$\mathcal{P}_{d=3}(x) = \frac{1}{2\sqrt{x}}, \quad 0 < x \leq 1. \tag{9}$$

Clearly, the observed overlap agrees qualitatively with the chaotic limit.

Another situation is shown in Fig. 7(a) where the data, a solid histogram, are taken for the ensemble which contains regular attractive pairing, $V_0 = -1$, plus uniformly random interactions in all channels with $L > 0$. The presence of pairing generates a significant probability for the paired ground state, a peak at $x = 1$. However there exists an "antipaired" state with $J = 0$ but a vanishingly small probability to be a ground state so that the space of the candidates for the ground state position is effectively two-dimensional, and we have

$$\mathcal{P}_{d=2}(x) = \frac{1}{\pi\sqrt{x(1-x)}}, \quad 0 < x \leq 1, \tag{10}$$

which again agrees qualitatively with the data, Fig. 7(a).

## 3. Conclusion

The regularities in the spectra of angular momentum conserving random two-body interactions in systems of $N$ fermions on a single-$j$ level are the result of geometric chaoticity, the inherent complexity of the $N$-body spin-$J$ states from the coupling of spin-$j$ particles. Treating the system as a Fermi gas in statistical equilibrium explained the trends both in the ground state spins and the yrast line. The role of pairing is minimal. The shell structure evident in the single particle occupancies sheds light on the limitations of the theory. This work was supported by the NSF grants 96-05207 and 00-70911.## REFERENCES

[1] C.W. Johnson, G.F. Bertsch and D.J. Dean, Phys. Rev. Lett. **80**, 2749 (1998).
[2] C.W. Johnson, G.F. Bertsch, D.J. Dean and I. Talmi, Phys. Rev. C **61**, 014311 (2000).
[3] D. Mulhall, V. Zelevinsky, and A. Volya Phys. Rev. Lett. **85**, 4016 (2000).